\documentclass[twocolumn,english,reprint, longbibliography, superscriptaddress, breaklinks=true, showkeys, showpacs=false, nofootinbib]{revtex4}
\usepackage[T1]{fontenc}
\usepackage[latin9]{inputenc}
\usepackage{color}
\usepackage{babel}
\usepackage{amsmath}
\usepackage{amssymb}
\usepackage{subfigure}
\usepackage{graphicx}
\usepackage{physics}
\usepackage[unicode=true,pdfusetitle,
 bookmarks=true,bookmarksnumbered=false,bookmarksopen=false,breaklinks=true,pdfborder={0 0 0},backref=false,colorlinks=true]
 {hyperref} 
\makeatletter
\@ifundefined{textcolor}{}
{%
 \definecolor{BLACK}{gray}{0}
 \definecolor{WHITE}{gray}{1}
 \definecolor{RED}{rgb}{1,0,0}
 \definecolor{GREEN}{rgb}{0,1,0}
 \definecolor{BLUE}{rgb}{0,0,1}
 \definecolor{CYAN}{cmyk}{1,0,0,0}
 \definecolor{MAGENTA}{cmyk}{0,1,0,0}
 \definecolor{YELLOW}{cmyk}{0,0,1,0}
}
\pdfoutput=1
\hypersetup{colorlinks=true,citecolor=blue,linkcolor=cyan,urlcolor=blue,filecolor= green, breaklinks=true}
\usepackage{url}
\usepackage{breakurl}
\makeatother

\begin{document}

\title{Operational connection  between predictability and entanglement in entanglement swapping from partially entangled pure states}

\author{Marcos L. W. Basso}
\email{marcoslwbasso@hotmail.com}
\address{Centro de Ci\^encias Naturais e Humanas, Universidade Federal do ABC, Avenida dos Estados 5001, Santo Andr\'e, S\~ao Paulo, 09210-580, Brazil}

\author{Jonas Maziero}
\email{jonas.maziero@ufsm.br}
\address{Departamento de F\'isica, Centro de Ci\^encias Naturais e Exatas, Universidade Federal de Santa Maria, Avenida Roraima 1000, Santa Maria, Rio Grande do Sul, 97105-900, Brazil}

\selectlanguage{english}%

\begin{abstract}
Complementarity and entanglement are fundamental features of Quantum Mechanics that were recently related in triality equalities that involve quantum coherence, the wave aspect of a qubit, and quantum predictability and quantum entanglement, the particle aspect of a qubit. In this article, we give an operational connection between predictability and entanglement in entanglement swapping from initially partially entangled states. For this, we show that the predictability of the pre-measurement one-qubit density matrix is directly related to the probability of the partially entangled component of the post-measurement state. Going even further, we analyze the entanglement swapping for partially entangled states in the light of complementarity relations and show that, in the cases where the entanglement increases after a Bell-basis measurement, the predictability is consumed when compared to the initially prepared state.

\end{abstract}


\maketitle

It is undeniable that the development of the field of Quantum Information has led to an ever increasing interest in the foundations of Quantum Mechanics (QM), or Quantum Foundations (QF). Since the beginning of QM, one of the pillars of QF is the famous Bohr's complementarity principle \cite{Bohr}, in which the wave-particle duality emerged as the main example.
Ever since, countless paths have been traced for quantifying the wave-particle properties of a quantum system (quantons) in terms of an elegant complementarity relation between predictability and visibility \cite{Wootters, Yasin, Engle, Ribeiro, Bera, Bagan, Coles, Hillery, Maziero, Leopoldo}. The generalization of such complementarity relation for $d$-dimensional quantons (qudits) came with the realization that the quantum coherence \cite{Baumgratz} is the natural generalization for the visibility of an interference pattern \cite{Bera, Bagan, Mishra}. As well, inspired by the mathematical development of QM and its consequent axiomatization, the authors in Ref. \cite{Maziero}, by exploring the properties of the density matrix of a quanton $A$, derived several complementarity relations between different measures of quantum coherence and predictability, with both satisfying the criteria established in Refs. \cite{Durr, Englert} for bone-fide measures of visibility and predictability. 

However, to fully characterize a quanton, it is not enough to consider its wave-particle aspect, one also has to account its correlations with other systems. For this, triality relations, also known as complete complementarity relations (CCRs), involving predictability, visibility and entanglement, were first suggested by Jakob and Bergou \cite{Janos} provided that a bipartite (or multipartite) quantum system $AB$ is in a  pure state. In contrast, by taking the purity of the bipartite quantum system as the main hypothesis, the authors in Ref. \cite{Marcos} derived CCRs of the type
\begin{align}
    C_{re}(\rho_A) + P_{vn}(\rho_A) + S_{vn}(\rho_A) = \log_2 d_A, \label{eq:ccr1}
\end{align}
where $C_{re}(\rho_A) := S_{vn}(\rho_{A diag}) - S_{vn}(\rho_{A})$ is the relative entropy of quantum coherence, with $\rho_{A diag}$ being the diagonal part of $\rho_{A}$, $P_{vn}(\rho_A) = \log_{2}d_{A} - S_{vn}(\rho_{A diag})$ is the corresponding bone-fide predictability measure and $S_{vn}(\rho_A) = - \Tr \rho_A \log_2 \rho_A$ is the von Neumann entropy, which is an entanglement monotone.
Besides, the first order approximation of Eq.  (\ref{eq:ccr1}) was also obtained in Ref. \cite{Marcos}, and it is the CCR based on the linear entropy: $C_{hs}(\rho_A) + P_{l}(\rho_A) + S_{l}(\rho_A) = (d_A - 1)/d_A$, where $C_{hs}(\rho_A)$ is Hilbert-Schmidt quantum coherence \cite{Baumgratz}, $P_{l}(\rho_A) := (d_{A}-1)/d_{A} - S_{l}(\rho_{A diag})$ and  $S_l(\rho_A) = 1 - \Tr \rho^2_A$. Besides, the predictability measure together with an entanglement monotone can be considered as a measure of path distinguishability in an interferometer, as recently discussed in Refs. \cite{Qureshi, Wayhs}. As well, the CCR given by Eq. (\ref{eq:ccr1}) has several interesting properties that were  explored recently in the literature: it is intrinsically connected to the notion of realism defined by Bilobran and Angelo \cite{Bilobran}, as shown in Ref. \cite{Jonas}; together with the CCR based on the linear entropy, it is invariant by global unitary operations, which implies that it is preserved under unitary evolution and it is invariant under Lorentz transformations \cite{CCRin}, which by its turn made possible the extension of the validity of the CCRs to curved spacetimes \cite{CCRcst}. 

Another remarkable aspect of Eq. (\ref{eq:ccr1}) is that both quantum coherence and quantum entanglement are properties of quantons that can be used as resources for tasks in Quantum Information and Quantum Computation. This motivated us to put forward a quantum resource theory of predictability \cite{Predres}, by relating the predictability of a given state with reference to an observable $X$ with quantum coherence with reference to observables mutually unbiased (MU) to $X$ of the state obtained from a non-revealing von Neumann measurement of $X$ as well as identifying its free states, free operations and resource states together with resource measures, in which $P_{vn}(\rho_A)$ and $P_l(\rho_A)$ are examples and can be seen as a way of quantifying how much the probability distribution expressed by the diagonal elements of $\rho_A$ differs from the uniform probability distribution.  In this manuscript, we give another step in this direction and report an operational connection between predictability and entanglement using a well known protocol of quantum information and relating the predictability of the pre-measurement one-qubit state with the probability of obtaining the maximally entangled component of the post-measurement state. To do this we use the protocol of entanglement swapping from initially partially entangled pure states.

Entanglement swapping is an essential primitive for quantum communication where one starts with two pairs of entangled qubits and makes a Bell-basis measurement (BBM) on one qubit from each pair, ending up maximally entangling the other two qubits, that possibly have never interacted directly \cite{Anton}. Entanglement swapping from initially partially entangled states was first studied in Ref. \cite{Bose}, for a particular class of initial states. In this article we make a thorough analysis by studying a more general class of initially partially entangled states and by showing that there is a plethora of partially entangled states as well as maximally entangled states that can be obtained after a BBM. Besides, for the partially entangled states obtained after a BBM, it is possible, by local operations, to purify this entanglement into a smaller number of maximally entangled pairs of quantons \cite{Schumacher}. This, as we will show, lead to a naturally operational connection between predictability and entanglement. Going even further, we analyze the entanglement swapping for partially entangled states in the light of complementarity relations and show that, in the cases where the entanglement increases after the BBM, the predictability is consumed when compared to the initially prepared state. Moreover, the triality relations will always remain valid before and after the BBM, which means the we have a interplay between predictability and entanglement before and after the BBM process.

Let us consider three laboratories handled by Alice, Bob, and Charlie. In addition, let us consider that Darwin, in another laboratory, prepares two pairs of qubits in a partially entangled pure state:
\begin{align}
   &  |\xi\rangle_{AC} = \sqrt{p}|00\rangle_{AC} + \sqrt{1-p}|11\rangle_{AC}, \\
    &  |\eta\rangle_{C'B} = \sqrt{q}|00\rangle_{C'B} + \sqrt{1-q}|11\rangle_{C'B},
\end{align}
with $p,q\in[0,1]$. The quantons $C$ and $C'$ are sent to Charlie, who makes a selective Bell basis measurement (BBM) on them. The quanton $A$ is sent to Alice and the quanton $B$ is sent to Bob. It is worth mentioning here that, in Ref. \cite{Bose}, the authors studied entanglement swapping of partially entangled pure states only for $p = q$. Therefore, the analysis of entanglement swapping of partially entangled pure states made in this manuscript is more general. The composed state of the four qubits can be written as
\begin{align}
    & \sqrt{2}|\xi\rangle_{AC}|\eta\rangle_{C'B} = \\ 
    & |\Phi_{+}\rangle_{CC'}\big(\sqrt{pq}|00\rangle_{AB}+\sqrt{(1-p)(1-q)}|11\rangle_{AB}\big) + \nonumber \\
    & |\Phi_{-}\rangle_{CC'}\big(\sqrt{pq}|00\rangle_{AB}-\sqrt{(1-p)(1-q)}|11\rangle_{AB}\big) + \nonumber\\
& |\Psi_{+}\rangle_{CC'}\big(\sqrt{p(1-q)}|01\rangle_{AB}+\sqrt{(1-p)q}|10\rangle_{AB}\big) + \nonumber \\ 
& |\Psi_{-}\rangle_{CC'}\big(\sqrt{p(1-q)}|01\rangle_{AB}-\sqrt{(1-p)q}|10\rangle_{AB}\big) \nonumber.
\end{align}

After Charlie makes a BBM, the possible post-measurement states are given by
\begin{align}
   & \ket{\phi_{\pm}}_{AB} = \frac{1}{N_{\phi}}\big(\sqrt{pq}|00\rangle_{AB}\pm\sqrt{(1-p)(1-q)}|11\rangle_{AB}\big), \label{eq:phipm}\\
   & \ket{\psi_{\pm}}_{AB} = \frac{1}{N_{\psi}}  \big(\sqrt{p(1-q)}|01\rangle_{AB}\pm\sqrt{(1-p)q}|10\rangle_{AB}\big), \label{eq:psipm}
\end{align}
with $N_{\phi} = \sqrt{pq+(1-p)(1-q)}$ and $N_{\psi} = \sqrt{p(1-q)+(1-p)q}$. The states $\ket{\phi_{\pm}}_{AB}$ and $\ket{\psi_{\pm}}_{AB}$ are obtained with probabilities:
\begin{align}
    & Pr(\phi_{\pm}) = \frac{1}{2}\Big(pq + (1-p)(1-q)\Big), \\
    & Pr(\psi_{\pm}) = \frac{1}{2}\Big(p(1-q) + (1-p)q\Big),
\end{align}
which are the same as the probabilities for Charlie to obtain the Bell states $\ket{\Phi_{\pm}}_{CC'}$ and $\ket{\Psi_{\pm}}_{CC'}$ respectively. Besides, it is easy to see that if we set $p=q=1/2$, the probability of any of the BBM outcomes is $1/4$ and the quantons $A$ and $B$ end up in a Bell (maximally entangled) state. As for the general case, we have that the von Neumann entropy of the local reduced state, $\rho^{\phi}_A=Tr_{B}(|\phi_{\pm}\rangle_{AB}\langle\phi_{\pm}|)$ and $\rho^{\psi}_A=Tr_{B}(|\psi_{\pm}\rangle_{AB}\langle\psi_{\pm}|)$,
is an entanglement monotone given, respectively, by $S_{vn}(\rho^{\phi}_A) = - a \log_2 a - b \log_2 b$ and $S_{vn}(\rho^{\psi}_A) = - c \log_2 c - d \log_2 d$, where $a = \frac{pq}{N_{\phi}}$, $b = \frac{(1-p)(1-q)}{N_{\phi}} $, $c = \frac{(1-p)q}{N_{\psi}} $, and $d = \frac{p(1-q)}{N_{\psi}} $. 

\begin{figure}[t]
    \centering
    \subfigure[$S_{vn}(\rho^{\phi}_A)$ as a function of $p$ for different values of $q$.]{{\includegraphics[scale = 0.45]{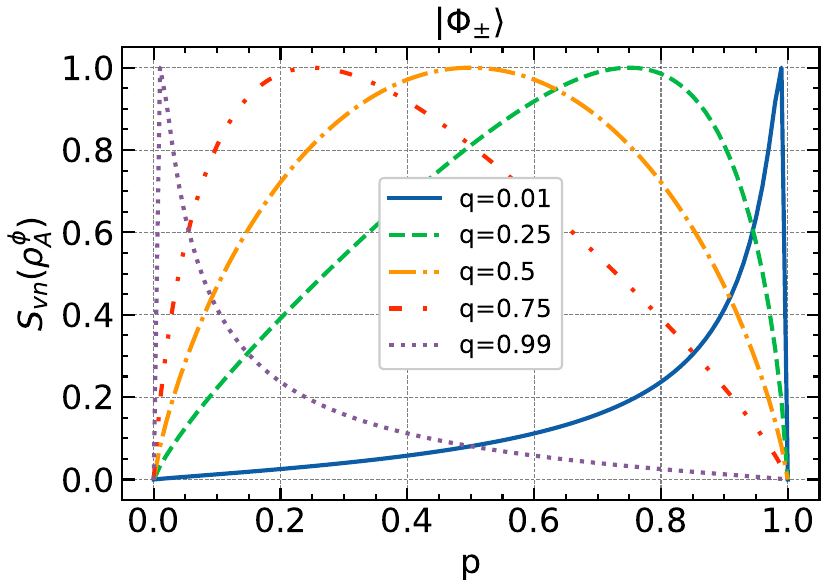}{\label{fig:a}} }}
    \subfigure[$S_{vn}(\rho^{\psi}_A)$ as a function of $p$ for different values of $q$.]{{\includegraphics[scale = 0.45]{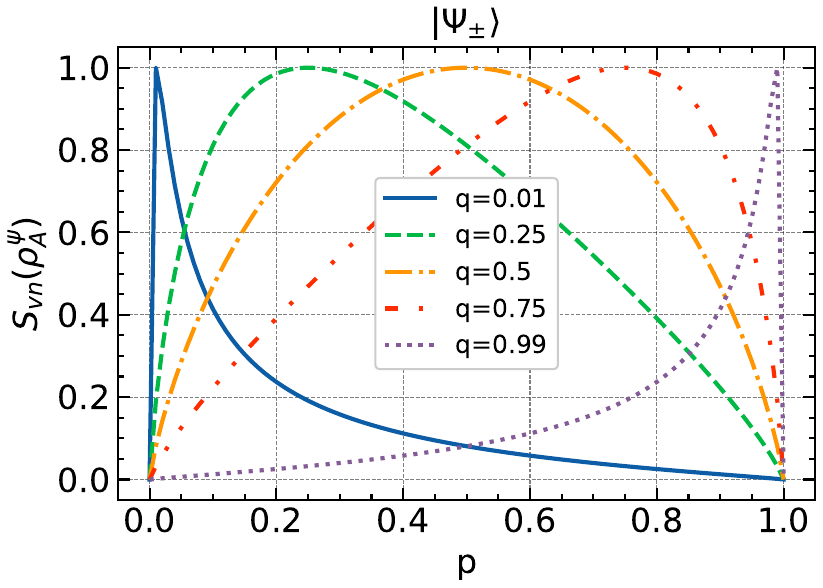}{\label{fig:b}} }}
  \caption{The von Neumann entropy of local reduces states $\rho^{\phi}_A, \rho^{\psi}_A$ as function of $p$ for different values of $q$.}    
\end{figure}

In Figs. \ref{fig:a} and \ref{fig:b}, are shown $S_{vn}(\rho^{\phi}_A)$  and $S_{vn}(\rho^{\psi}_A)$ as a function of $p$ for different values of $q$. Thus one can see that the post measurement states $\ket{\phi_{\pm}}_{AB}$, $\ket{\psi_{\pm}}_{AB}$ are entangled in general, except when $q = 0$ or $1$, and similarly for $p$, which corresponds to the trivial case where $|\xi\rangle_{AC}, |\eta\rangle_{C'B}$ are separable. Besides, further analysis shows that, for any given $q$, $S_{vn}(\rho^{\phi}_A)$ and $S_{vn}(\rho^{\psi}_A)$ vary continuously between $0$ and $1$ and then go back to $0$ as $p$ varies in the interval $[0,1]$. This means that, for any given $q$, there always exists a value of $p$ such that $S_{vn}(\rho^{\phi}_A)$  and $S_{vn}(\rho^{\psi}_A)$ are maximal. Indeed, it is easy to show this by looking at the first and second derivatives of $S_{vn}(\rho^{\phi}_A)$  and $S_{vn}(\rho^{\psi}_A)$ in relation of $p$:
\begin{align}
& \frac{d}{dp}S_{vn}(\rho^{\phi}_A) = 0 \ \land \  \frac{d^2}{dp^2}S_{vn}(\rho^{\phi}_A) < 0 \iff p = 1 - q, \nonumber \\
& \frac{d}{dp}S_{vn}(\rho^{\psi}_A) = 0 \ \land \  \frac{d^2}{dp^2}S_{vn}(\rho^{\psi}_A) < 0 \iff p = q. \nonumber
\end{align}

Therefore, if one chooses $p = q$, as in Ref. \cite{Bose}, we have that, for a given $q$, $S_{vn}(\rho^{\psi}_A)$ reaches its maximum possible value while $S_{vn}(\rho^{\phi}_A)$ does not. However, we have the other possibility where one choose $p = 1 - q$ such that $S_{vn}(\rho^{\phi}_A)$ will reach its maximum possible value while $S_{vn}(\rho^{\psi}_A)$ will not. In addition, there is a plethora of cases where $p \neq q$ or $p \neq 1-q$ such that $\ket{\phi_{\pm}}_{AB}$ and $\ket{\psi_{\pm}}_{AB}$ will end up in a partially entangled state. For instance, if $p = 0.1$ and $q = 0.75$, the initial states $ |\xi\rangle_{AC} $ and $ |\eta\rangle_{C'D} $ have $S_{vn}(\rho^{\xi}_A) = S_{vn}(\rho^{\xi}_C) \approx 0.4689$ and $S_{vn}(\rho^{\eta}_{C'}) = S_{vn}(\rho^{\eta}_B) \approx 0.8112$, respectively. Meanwhile, Alice and Bob will obtain the post-measurement states $\ket{\phi_{\pm}}_{AB}$ and $\ket{\psi_{\pm}}_{AB}$ with $S_{vn}(\rho^{\phi}_A) = S_{vn}(\rho^{\phi}_B) \approx 0.8112$ with probability $Pr(\phi_{\pm}) = 0.15 $ and $S_{vn}(\rho^{\psi}_A) = S_{vn}(\rho^{\psi}_B) \approx 0.2222$ with $Pr(\psi_{\pm}) = 0.35 $. Thus, one can see that Alice and Bob obtain the post-measurement states with the lower entanglement more frequently. However, if there is a large enough ensemble of partially entangled quantons in the beginning of the protocol, it is possible, by local operations, to purify this entanglement into a smaller number of maximally entangled pairs of quantons.

\begin{figure}[t]
    \centering
    \subfigure[$Pr(\Phi_{\pm})$, $Pr(\Psi_{\pm})$ and $P_l(\rho_i)$ as a function of $q$, where $\rho_i$ is one of the initial local density matrices.]{{\includegraphics[scale = 0.35]{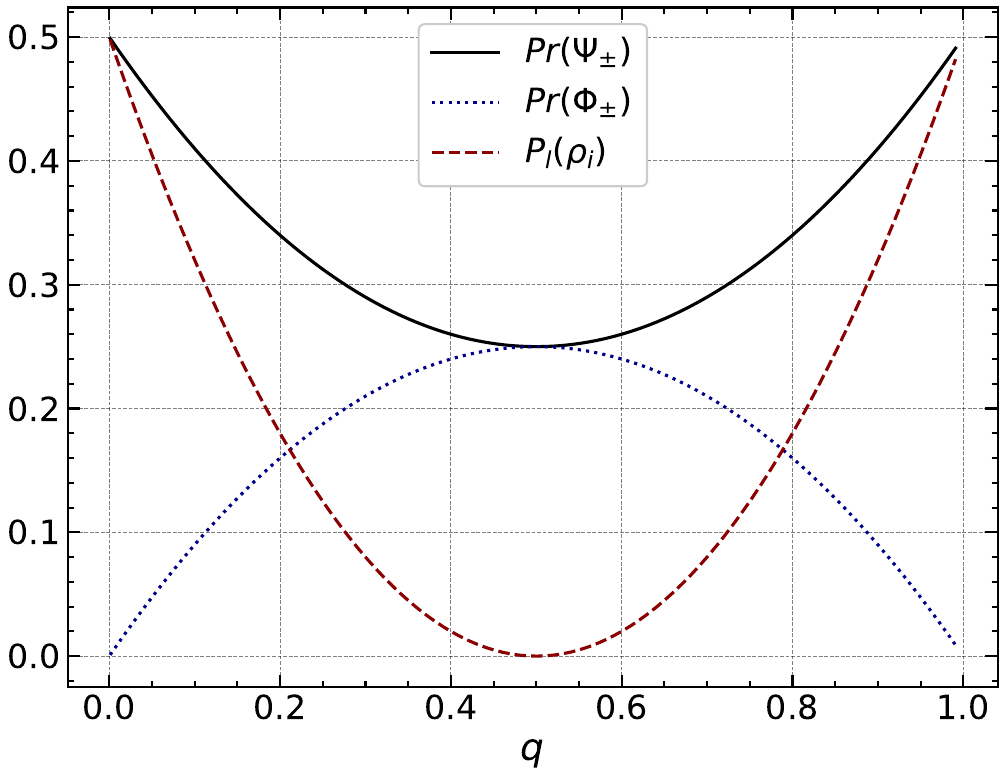}{\label{fig:prpl}} }}
    \subfigure[$S_{vn}(\rho^{i}_A)$, $S_{vn}(\rho^{f}_A)$, $P_{vn}(\rho^{i}_A)$, $P_{vn}(\rho^{f}_A)$ as a function of $q$.]{{\includegraphics[scale = 0.35]{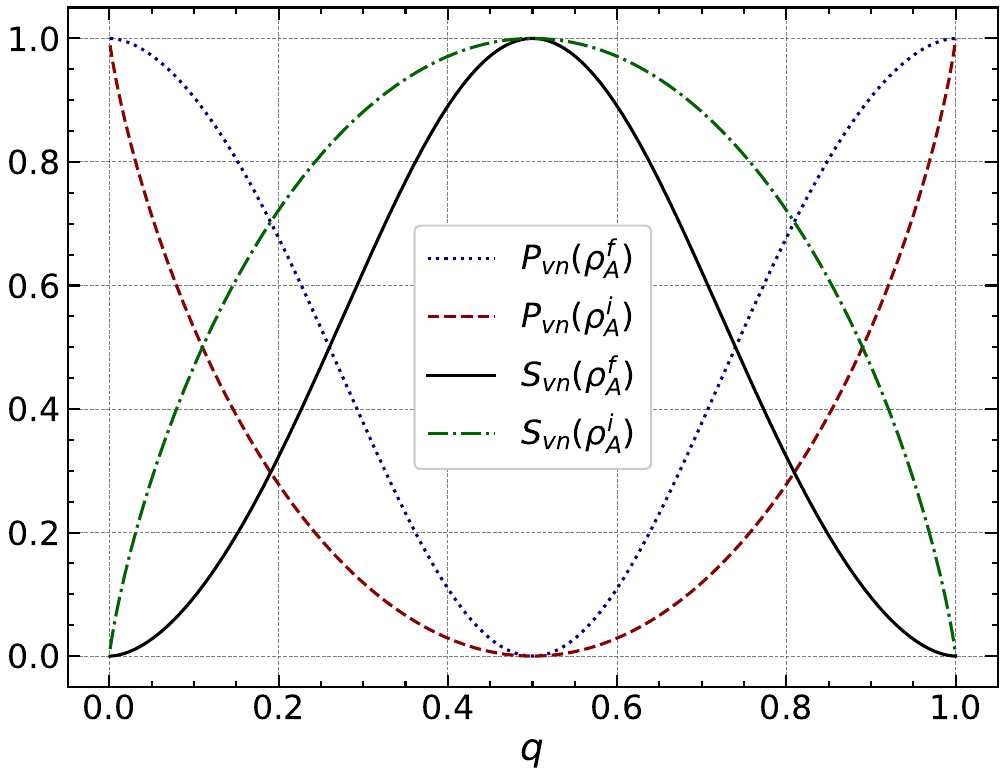}{\label{fig:c}} }}
  \caption{Interplay between the one-qubit predictability before a Bell-base measurement (BBM) and the entanglement and probability amplitudes of the state after the BBM.}    
\end{figure}

Now, for completeness, let us consider the  case specified by $p = 1 - q$, for which
\begin{align}
& |\xi\rangle_{AC}|\eta\rangle_{C'B}  = \nonumber \\ & \sqrt{(1-q)q}|\Phi_{+}\rangle_{CC'}|\Phi_{+}\rangle_{AB} + \sqrt{(1-q)q}|\Phi_{-}\rangle_{CC'}|\Phi_{-}\rangle_{AB} \nonumber \\
&  + |\Psi_{+}\rangle_{CC'}\frac{1}{\sqrt{2}}\big(\sqrt{(1-q)^{2}}|01\rangle_{AB}+\sqrt{q^{2}}|10\rangle_{AB}\big)\nonumber \\ & + |\Psi_{-}\rangle_{CC'}\frac{1}{\sqrt{2}}\big(\sqrt{(1-q)^{2}}|01\rangle_{AB}-\sqrt{q^{2}}|10\rangle_{AB}\big). \label{eq:statepq}
\end{align}
If Charlie makes a BBM, then he, as well as Alice and Bob, will end up with the state $\ket{\Phi_{\pm}}$ 
with the probability $Pr(\Phi_{\pm}) = q(1-q)$. In the other cases, Charlie will end up with states $\ket{\Psi_{\pm}}$ with probabilities $Pr(\Psi_{\pm}) = \frac{1}{2}((1-q)^2 + q^2)$ and Alice and Bob will share a partially entangled state given by $\ket{\psi_{\pm}}$ with $p = 1 - q$. For $q \approx 0$ or $q \approx 1$, one can see that $Pr(\Psi_{\pm}) >> Pr(\Phi_{\pm})$. For instance, for $q = 0.99$, $Pr(\Psi_{\pm}) = 0.4901, \ \ \ Pr(\Phi_{\pm}) =  0.0099$ while for $q = 0.75$, $Pr(\Psi_{\pm}) = 0.3125, \ \ \ Pr(\Phi_{\pm}) = 0.1875$. Actually, as one can see in Fig. \ref{fig:prpl}, $Pr(\Psi_{\pm}) \ge Pr(\Phi_{\pm})$ for all $q \in [0,1]$. However, there is a non-null probability that Alice and Bob will end up with a maximally entangled state even when the initial states are partially entangled. As already notice in Ref. \cite{Bose}, she actually increases the magnitude of entanglement she shares with Bob. However, the authors did not explain from where this extra entanglement came from. And one of the main goals of this article is to explain the origin of this extra entanglement. In the other cases, Alice and Bob will end up with lower entanglement states in comparison with the initial ones. However, as was noticed in Ref. \cite{Bose}, if there is a large enough ensemble of partially entangled quantons in the beginning of the protocol, it is possible, by local operations, to change the states of a certain fraction of the shared pairs of Alice and Bob to Bell states at the cost of decreasing the entanglement of the other shared pairs even further.


It is intriguing the fact that, for $p = q$ or $p = 1 -q$, there is a non-null probability of obtaining a maximally entangled state when we start with initial states that are partially (even almost not) entangled. However, this fact turns out to be very reasonable provided that we analyse it in the light of complementarity relations like the one given by Eq. (\ref{eq:ccr1}). In these cases, we have that all the pre-measurement local quantum systems $A, B, C$ and $C'$ have the same predictability and share the same entanglement, i.e.,
\begin{align}
    & S_{vn}(\rho^i_j) = - q \log_2 q - (1-q)\log_2(1-q), \\
    & P_{vn}(\rho^i_j) = 1 + q \log_2 q + (1-q)\log_2(1-q), \label{eq:first}
\end{align}
with $j =  A,B,C,C'$, such that $P_{vn}(\rho^i_j) + S_{vn}(\rho^i_j) = 1$. The index $i$ refers to the initial reduced density matrix before the BBM. After the BBM, at each run of the protocol, we will have a non-null probability of obtaining a maximally entangled state, for which
\begin{align}
P_{vn}(\rho^f_j) = 0, \ \ S_{vn}(\rho^f_j) = 1 \text{, for } j = A,B,C,C'.
\end{align}
This means that the predictability was consumed
for the generation of entanglement such that, after the BBM, Alice and Bob end up sharing a maximally entangled state, as well as does Charlie. Of course, besides increasing, the entanglement was redistributed between the quantons such that, after the BBM, $A$ and $B$ are entangled as well as are $C$ and $C'$. Therefore, such analysis remains valid since $P_{vn}$ and $S_{vn}$ are functions of a single local quanton, and during the protocol we always have a pair of bipartite pure quantum systems. 

However, in most of the cases, Alice and Bob will end up with a partially entangled state, which has a lower or equal degree of entanglement when compared with the initial states, while Charlie will end up with a maximally entangled state. This means that the predictability of $A$ and $B$ in the end has increased or remained the same, while the predictability of $C$ and $C'$ is always zero. For instance, if $p = 1 - q$, most of the times Charlie will obtain the state $\ket{\Psi_{\pm}}$ and Alice and Bob will share the state $\ket{\psi_{\pm}}$ such that $P_{vn}(\rho^{f}_A) \ge P_{vn}(\rho^i_A)$, as one can see in Fig. \ref{fig:c}. The same analysis can be done for $p = q$. For $p \neq q$ and $p \neq 1 - q$, the predictability and entanglement of the post measurement state obtained by Alice and Bob will be a function of $p$ and $q$, as one can see from Eqs. (\ref{eq:phipm}) and (\ref{eq:psipm}). However $P_{vn}(\rho_j) + S_{vn}(\rho_j) = 1$, for  $j = A,B,C,C'$, will remain valid before and after the BBM, which means the we have an interplay between predictability and entanglement before and after the BBM, once there's no coherence in the one-qubit states before and after the BBM. 

Beyond that, if $p = 1-q$ the predictability of the pre-measurement one-qubit density matrix is directly related to the probability of Charlie obtaining the states  $\ket{\Psi_{\pm}}$ and Alice and Bob sharing the state the states $\ket{\psi_{\pm}}$, i.e., by making a first order expansion of Eq.(\ref{eq:first}), it is possible to see that
\begin{align}
    & P_{vn}(\rho^i_j) = \frac{1}{\ln 2}(q^2 + (1 -q)^2) + \Big(1 - \frac{1}{\ln 2}\Big) + \mathcal{O}(q^3) \\
    & =  \frac{1}{\ln 2}(Pr(\Psi_+) + Pr(\Psi_-)) + \Big(1 -  \frac{1}{\ln 2}\Big) + \mathcal{O}(q^3), \nonumber 
\end{align}
for $j = A, B$, where $Pr(\Psi_{\pm}) = \frac{1}{2}(q^2 + (1 -q)^2)$ is the probability of Charlie ending up with the one of the states $\ket{\Psi_{\pm}}$ and we used the fact that $\ln x \simeq x - 1$. As pointed out in the introduction, the linear predictability $P_l$ can be seen as the first order expansion of $ P_{vn}$, which implies that we have the following exact result:
\begin{align}
    Pr(\Psi_{\pm}) = \frac{1}{2}\Big(\frac{1}{2}+P_l(\rho^i_j)\Big).
\end{align}
For the maximally entangled components of the state in Eq. (\ref{eq:statepq}), we have $Pr(\Phi_{\pm}) = \frac{1}{2}\Big(\frac{1}{2}-P_l(\rho^i_j)\Big)$.

For instance, if, initially, we have $P_l(\rho^i_j) = 0$, it means that $p = q = 1/2$ and the initial states are maximally entangled. Therefore, the probability of any of the BBM outcomes is $1/4$ and the quantons $A$ and $B$ end up in a Bell (maximally entangled) state. In the other cases, we can see that the probability of obtaining partially entangled states after BBM is directly proportional to the predictability of the initial one-qubit states, as one can see in Fig. \ref{fig:prpl}. This means that, in the context of the entanglement swapping protocol, we are able to relate the pre-measurement predictability with the entanglement obtained at the output. This connection can be expressed in operational terms as follows: \textit{the pre-measurement predictability is directly related to the probability of Alice and Bob obtaining partially entangled states after Charlie makes a BBM, which, by its turn, with a large enough copies, can be used for entanglement distillation. Otherwise, if the initial predictability is zero,  Alice and Bob will end up with a maximally entangled state and they will not need to purify their sample.} In other words, if, initially, Alice, Bob and Charlie have a large enough ensemble of quantum system prepared in the state given by Eq. (\ref{eq:statepq}) and, for instance, if such states have a high local predictability (which means that the entanglement is initially small), after Charlie makes a BBM on these states, Alice and Bob will end up with a large fraction of partially entangled states that, in turn, can be used for entanglement distillation, and therefore, a certain fraction of the shared pairs of Alice and Bob will be maximally entangled. The same interpretation holds for $p = q$.


To summarize, in this work, we were able to relate the predictability of the pre-measurement one-qubit state with the probability of obtaining the maximally entangled component of the post-measurement state. This connection was expressed in operational terms, i.e., we used the protocol of entanglement swapping from initially partially entangled states. To do this, we studied such protocol by considering a more general class of initially partially entangled states than was regarded in Ref. \cite{Bose}. We have found that there is a plethora of partially entangled states as well as maximally entangled states that can be obtained after a Bell-basis measurement (BBM). Besides, for the partially entangled states obtained after the BBM, it is possible, by local operations, to purify this entanglement into a smaller number of maximally entangled pairs of quantons. This, as we showed, led to a natural operational interpretation for predictability of the prepared states in terms of the probability of Alice and Bob obtaining partially entangled states after Charlie makes a BBM, which can be used for entanglement distillation. In contrast, if the initial predictability is zero,  Alice and Bob will end up with a maximally entangled state and they will not need to purify their qubits. In addition, we analyzed the entanglement swapping for partially entangled states in the light of complete complementarity relations and showed that, in the cases where the entanglement increases after the BBM, the predictability is consumed when compared to the initially prepared state. Moreover, $P_{vn}(\rho_j) + S_{vn}(\rho_j) = 1$ will always remain valid before and after the BBM, which means the we have a interplay between predictability and entanglement before and after the BBM. 

This work was supported by the Federal University of ABC (UFABC), process 23006.000123/2018-23, by the National Institute for the Science and Technology of Quantum Information (INCT-IQ), process 465469/2014-0, and by the National Council for Scientific and Technological Development (CNPq), process 309862/2021-3.

\end{document}